\documentclass[twocolumn,tighten,trackchanges]{aastex63}
\usepackage{mathtools}
\usepackage{txfonts} 
\usepackage{hyperref}
\usepackage{color}

\defcitealias{PaperI}{EHTC~I}
\defcitealias{PaperII}{EHTC~II}
\defcitealias{PaperIII}{EHTC~III}
\defcitealias{PaperIV}{EHTC~IV}
\defcitealias{PaperV}{EHTC~V}
\defcitealias{PaperVI}{EHTC~VI}
\defcitealias{PaperVII}{EHTC~VII}
\defcitealias{PaperVIII}{EHTC~VIII}
\defcitealias{Narayan_2021}{N21}
\defcitealias{Gelles_2021}{G21}

\newcommand{\npix}{$N_{\rm pix}$}%
\newcommand{\m}{M\,87*}
\newcommand{\s}{Sgr\,A*}
\begin{document}               

\turnoffeditone

\title{
Bayesian Accretion Modeling: Axisymmetric Equatorial Emission in the Kerr Spacetime
}
\shorttitle{Bayesian Accretion Modeling}

\author[0000-0002-7179-3816]{Daniel~C.~M~Palumbo}
\email{daniel.palumbo@cfa.harvard.edu}
\affiliation{Center for Astrophysics $\vert$ Harvard \& Smithsonian, 60 Garden Street, Cambridge, MA 02138, USA}
\affiliation{Black Hole Initiative at Harvard University, 20 Garden Street, Cambridge, MA 02138, USA}

\author[0000-0001-8053-4392]{Zachary~Gelles}
\affiliation{Center for Astrophysics $\vert$ Harvard \& Smithsonian, 60 Garden Street, Cambridge, MA 02138, USA}
\affiliation{Black Hole Initiative at Harvard University, 20 Garden Street, Cambridge, MA 02138, USA}

\author[0000-0003-3826-5648]{Paul Tiede}
\affiliation{Center for Astrophysics $|$ Harvard \& Smithsonian, 60 Garden Street, Cambridge, MA 02138, USA}
\affiliation{Black Hole Initiative at Harvard University, 20 Garden Street, Cambridge, MA 02138, USA}t

\author[0000-0001-9939-5257]{Dominic O. Chang}
\affiliation{Black Hole Initiative at Harvard University, 20 Garden Street, Cambridge, MA 02138, USA}
\affiliation{Center for Astrophysics $|$ Harvard \& Smithsonian, 60 Garden Street, Cambridge, MA 02138, USA}

\author[0000-0002-5278-9221]{Dominic~W.~Pesce}
\affiliation{Center for Astrophysics $\vert$ Harvard \& Smithsonian, 60 Garden Street, Cambridge, MA 02138, USA}
\affiliation{Black Hole Initiative at Harvard University, 20 Garden Street, Cambridge, MA 02138, USA}

\author[0000-0003-2966-6220]{Andrew Chael}
\affiliation{Princeton Gravity Initiative, Jadwin Hall, Princeton University, Princeton NJ 08544, USA}

\author[0000-0002-4120-3029]{Michael~D.~Johnson}
\affiliation{Center for Astrophysics $\vert$ Harvard \& Smithsonian, 60 Garden Street, Cambridge, MA 02138, USA}
\affiliation{Black Hole Initiative at Harvard University, 20 Garden Street, Cambridge, MA 02138, USA}

\begin{abstract}
The Event Horizon Telescope (EHT) has produced images of two supermassive black holes, Messier~87* (\m{}) and Sagittarius~A* (\s{}). The EHT collaboration used these images to indirectly constrain black hole parameters by calibrating measurements of the sky-plane emission morphology to images \edit1{of } general relativistic magnetohydrodynamic (GRMHD) simulations.  Here, we develop a model for directly constraining the black hole mass, spin, and inclination through signatures of lensing, redshift, and frame dragging, while simultaneously marginalizing over the unknown accretion and emission properties. By assuming optically thin, axisymmetric, equatorial emission near the black hole, our model gains orders of magnitude in speed over similar approaches that require radiative transfer. Using 2017 EHT \m{} baseline coverage, we use fits of the model to itself to show that the data are insufficient to demonstrate existence of the photon ring. We then survey time-averaged GRMHD simulations fitting EHT-like data, and find that our model is best-suited to fitting magnetically arrested disks, which are the favored class of simulations for both \m{} and \s{}. For these simulations, the best-fit model parameters are within ${\sim}10\%$ of the true mass and within ${\sim}10^\circ$ for inclination. With 2017 EHT coverage and 1\% fractional uncertainty on amplitudes, spin is unconstrained. Accurate inference of spin axis position angle depends strongly on spin and electron temperature. Our results show the promise of directly constraining black hole spacetimes with interferometric data, but they also show that nearly identical images permit large differences in black hole properties, highlighting
\edit1{degeneracies between the plasma properties, spacetime, and most crucially, the unknown emission geometry when studying lensed accretion flow images at a single frequency.}

\end{abstract}


\section{Introduction}

In April 2017, the Event Horizon Telescope (EHT) observed the supermassive black hole at the center of the giant elliptical galaxy Messier~87. The EHT image of this black hole (hereafter \m) showed a deep brightness depression indicative of the shadow of a black hole with a mass near $6.5\times10^9$ $M_\odot$ \citep[][hereafter EHTC I-VI]{PaperI, PaperII, PaperIII, PaperIV, PaperV, PaperVI}. The moderate brightness asymmetry across the ring-like structure suggested Doppler effects from rotation about a nearly face-on orbital axis. Polarized images and analyses have been presented in \citet[][hereafter EHTCVII-VIII]{PaperVII, PaperVIII}. More recently, the EHT released images and mass measurements for the supermassive black hole at our galactic center, \edit1{Sagittarius A* (\s{})}  \citep{SgrA_PaperI,SgrA_PaperII,SgrA_PaperIII,Sgra_PaperIV,SgrA_PaperV,SgrA_PaperVI}.

In the total intensity and polarized analyses of both \m{} and \s{} so far performed by the EHT, models of the sky intensity distribution are fit to interferometric visibilities with no assumptions regarding the underlying physics besides positivity of the observed intensity. This approach is ideal for representing what the EHT saw on the sky, but it is not naturally suited for measurements of physical parameters such as black hole mass or spin.

\begin{figure*}
    \centering
    \includegraphics[height=8cm]{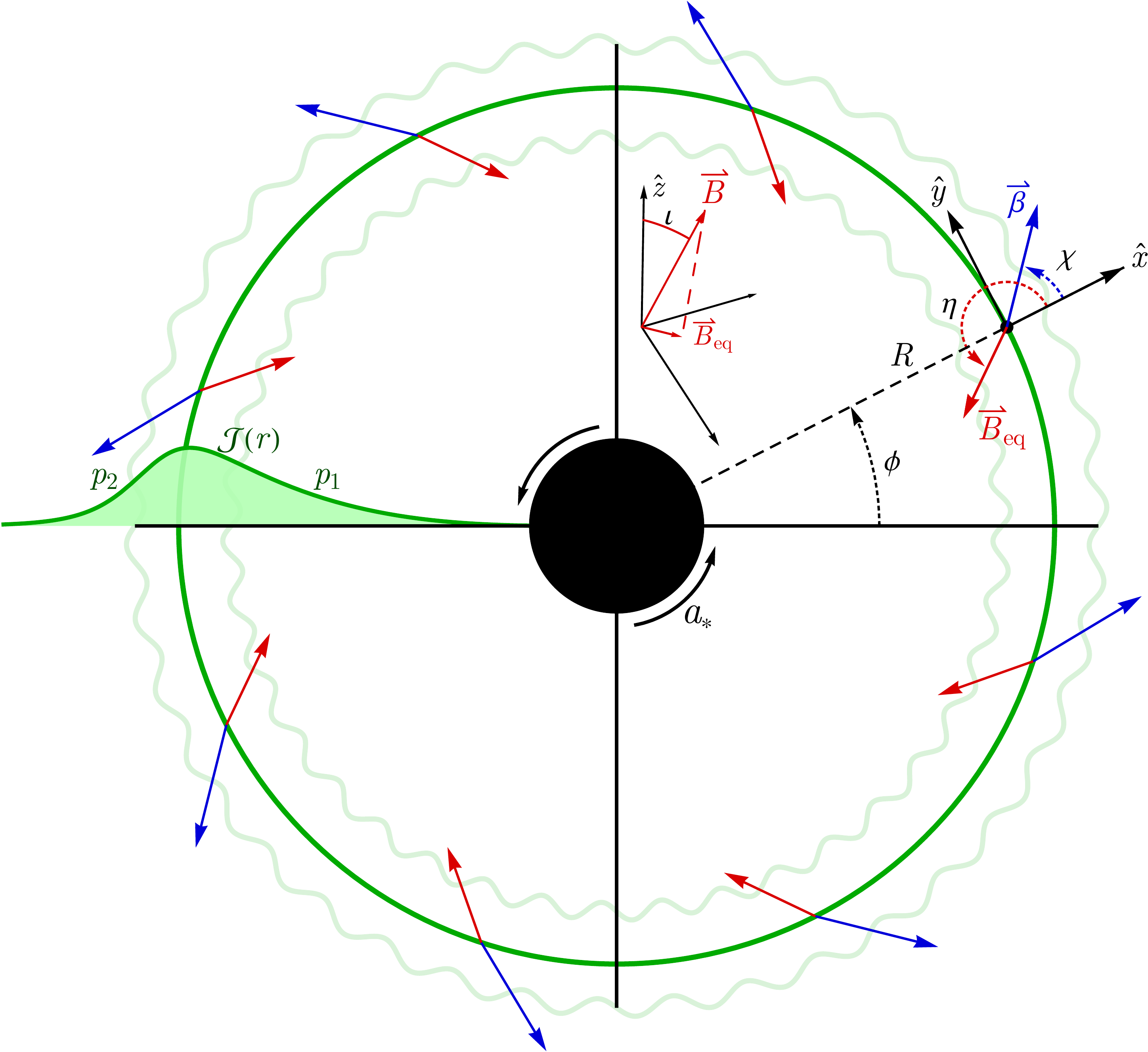}\hfill
    \includegraphics[height=8cm]{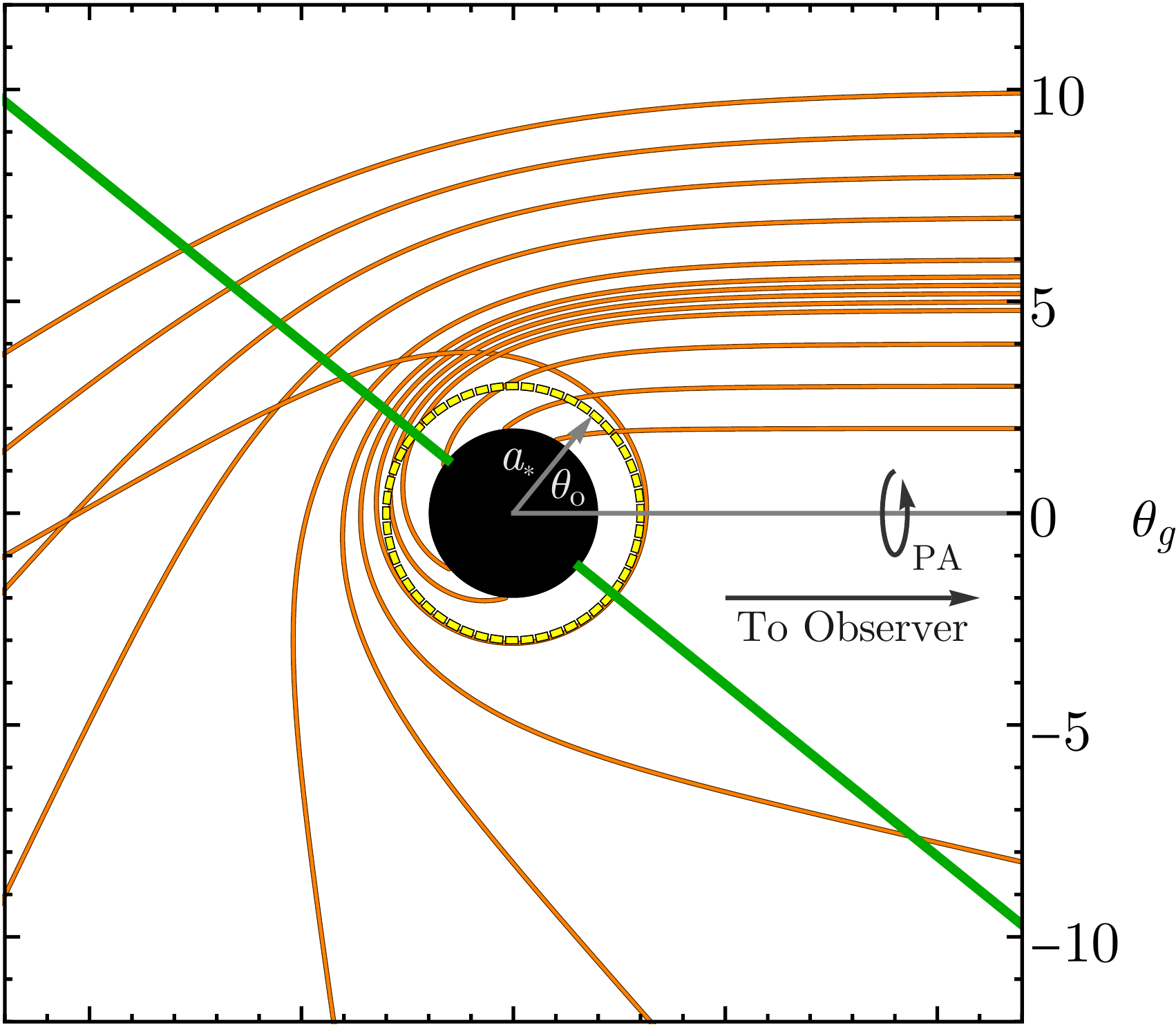}
    \caption{Summary of the equatorial emission model. Left: Emission model viewed from above, which is defined by an axisymmetric magnetic field ($\vec{B}$) and local fluid velocity ($\vec{\beta}$) that are given with respect to the local ZAMO tetrad. The magnetic field direction is determined by a pair of spherical angles $(\eta, \iota)$, while the velocity vector lies in the equatorial plane with a magnitude $\beta$ and angle $\chi$ with respect to the local radial direction. 
    The emissivity has a radial profile $\mathcal{J}(r)$, which we typically model as a double power law defined by a characteristic emission radius $R$ and a pair of indices $(p_1, p_2)$. Right: Side view showing null geodesics that reach a distant observer at viewing inclination $\theta_{\rm o}$ with respect to the angular momentum axis of the black hole and accretion flow. The black hole parameters $(\theta_{\rm g}, a_\ast)$ affect the image through gravitational lensing, redshift, and frame dragging, while the positional angle (PA) corresponds to a trivial image rotation. For simplicity, the geodesics shown correspond to a Schwarzschild black hole ($a_\ast = 0$). 
    }
    \label{fig:pedagogy}
\end{figure*}

As discussed extensively in \citetalias{PaperV} and \citetalias{PaperVI}, the mass measurements resulting from ring features on the sky require a calibration against theoretical predictions, in this case general relativistic magnetohydrodynamic (GRMHD) simulations, to account for the effect of differing emission geometries upon the resulting ring size. However, some criticisms of the EHT results center on the confounding potential of the detailed plasma configuration near the black hole \citep{Gralla_2019_photonrings}. Emission geometries that are not typically produced by GRMHD (such as emissivity distributions that truncate far from the horizon) would alter the calibration. Moreover, because of the computational expense of GRMHD simulations, the accretion and emission parameter space is only sparsely sampled and crudely characterized.

An alternative approach is to fit the black hole spacetime and emissivity distribution directly, potentially allowing for more general emission geometries and no longer relying upon GRMHD calibration. In particular, a series of papers \citep{Broderick_2009_RIAF, Broderick_2011_RIAF, Broderick_2014_RIAF, Broderick_2016, Pu_2016, Pu_2018} has developed semi-analytic model fitting implementations of radiatively inefficient accretion flows (RIAFs), which have been used to model the \m{} and \s{} accretion flows for decades \citep[see, e.g.][]{Ichimaru1977, Rees_1982,Narayan_1994,Narayan_1995,Reynolds_1996}. \edit1{Recent efforts have imposed additional conservation laws on the plasma profiles of RIAFs in Kerr and non-Kerr spacetimes \citep{Ozel_2021,Younsi_2021}. These results find characteristic emission radii comparable to most GRMHD simulations.}


These models can directly constrain black hole and accretion parameters using very-long baseline interferometry (VLBI) measurements in a Bayesian modeling framework. More recently, \citet{Tiede_2020} expanded these models to include time variability, demonstrating precise and accurate posterior estimation of mass and spin in self-fits of a physical model of infalling hotspots. However, these approaches have been limited by their severe computational expense, which involves numerical integration of radiative transfer along each ray to compute an image at each likelihood evaluation.

In this paper, we present a new emissivity model fitting paradigm: \texttt{KerrBAM}, for Kerr Bayesian Accretion Modeling. This approach develops a toy model introduced by \citet{Narayan_2021} (hereafter \citetalias{Narayan_2021}) and expanded to Kerr by \citet{Gelles_2021} (hereafter \citetalias{Gelles_2021}). Our model assumes emission that is optically thin, axisymmetric, and equatorial. These assumptions eliminate the need for numerical radiative transfer, allowing semi-analytic computation of model images that gain orders of magnitude in speed relative to previous modeling approaches. 
\texttt{KerrBAM} allows fitting of black hole parameters such as mass, spin, viewing inclination, and spin axis position angle while marginalizing over the unknown accretion and emission properties.

We introduce our approach to modeling a black hole accretion system and show image-domain comparisons to GRMHD simulations in \autoref{sec:formalism}. Example fits to synthetic VLBI data are shown in \autoref{sec:fits}. We conclude with a discussion in \autoref{sec:discussion}.

\section{Modeling Paradigm}
\label{sec:formalism}

Our goal in modeling the accretion flow is to be able to marginalize over the morphology and physical details of the emitting plasma when fitting interferometric data, producing  measurements with conservative, physically motivated uncertainties of black hole parameters of interest: the mass-to-distance ratio, spin, viewing inclination, and position angle of the projected spin axis. We approach this problem with a model for synchrotron emission from an orbiting plasma, specified everywhere in space along the Kerr midplane. Though the synchrotron emissivity predicts the polarized image, in this article we use only the predicted Stokes $I$ total intensity image.

\begin{table}[t]
\centering
\caption{
Model parameters and prior ranges used for fitting simulations of \m. 
}
\label{tab:params}
\begin{tabular}{ccc}
\hline
\hline
\textbf{Ray Tracing Parameters} & \textbf{Symbol} & \textbf{Prior Range} \\
\hline
Angular Gravitational Radius ($\mu$as) & $\theta_{\rm g}$ & $[1.5,4.5]$\\
\hline
Dimensionless Spin & $a_*$ & [-1,0]\\
\hline
Observer Inclination & $\theta_{\rm o}$ & $[0,40^\circ]$\\
\hline
Field of View ($\mu$as along edge) & FOV & 80 (Fixed)\\
\hline
Number of Pixels (along edge)  & \npix & 120 (Fixed)\\
\hline
Maximum photon winding number & $n_{\rm max}$ & 1 (Fixed)\\
\hline
\hline
\textbf{Fluid Parameters} & & \\
\hline
Fluid Speed (fraction of $c$) & $\beta$ & $[0,0.9]$\\
\hline
Equatorial Fluid Velocity Angle & $\chi$ &  $[-\pi,0]$\\
\hline
Equatorial Magnetic Field Angle & $\eta$ & $\chi+\pi$\\
\hline
Vertical Magnetic Field Angle & $\iota$ & $[0,\frac{\pi}{2}]$ \\
\hline
Spectral Index & $\alpha_\nu$ & $[-3,3]$\\
\hline
Synchrotron Cross Product Index & $\alpha_\zeta$ & $[-3,3]$\\
\hline
\hline
\textbf{Emission Profile Parameters} & &\\
\hline
Characteristic Radius (M) & $R$ & $[1,8]$ \\
\hline
Inner Index & $p_1$  & $[0.1,10]$ \\
\hline
Outer Index & $p_2$   & $[1,10]$ \\
\hline
\hline
\textbf{Observer Parameters} & &\\
\hline
Total Stokes $I$ Flux & $I_{\rm tot}$ & Fixed \\
\hline
Projected Spin Axis Position Angle & PA & $[243^\circ,333^\circ]$\\
\hline
\end{tabular}
\end{table}

The physics of the fluid model is exactly as in \citetalias{Narayan_2021} and \citetalias{Gelles_2021}; we specify fluid motion with respect to the zero angular momentum observer (ZAMO) and generate all observable quantities from spatially uniform physical parameters. In brief: we semi-analytically trace rays backwards from the observer screen to the Kerr midplane in Boyer-Lindquist coordinates, boost into the frame co-moving with the local fluid to evaluate the synchrotron emissivity, boost back out of the frame, and transport the polarization vector out along the geodesic. We avoid radiative transfer by assuming that the fluid is optically thin everywhere; the flat space radiative transfer equation for received emission from a path of length $s$ through material in the absence of absorption and scattering is then simply
\begin{align}
    I_{\nu}(s) = j_\nu s,
\end{align}
where $j_\nu$ is the emissivity \citep{R&L}. We specify our model as a disk of constant thickness with $j_\nu$ absorbing variation in physical quantities of the flow; thus, $s$ varies only relatively according to the equatorial component of the photon momentum. The model does not possess physical units for absolute quantities related to energy or volume, so we recast the emissivity as the unitless function $\mathcal{J}$.

\begin{figure*}
    \centering
    \includegraphics[width=\textwidth]{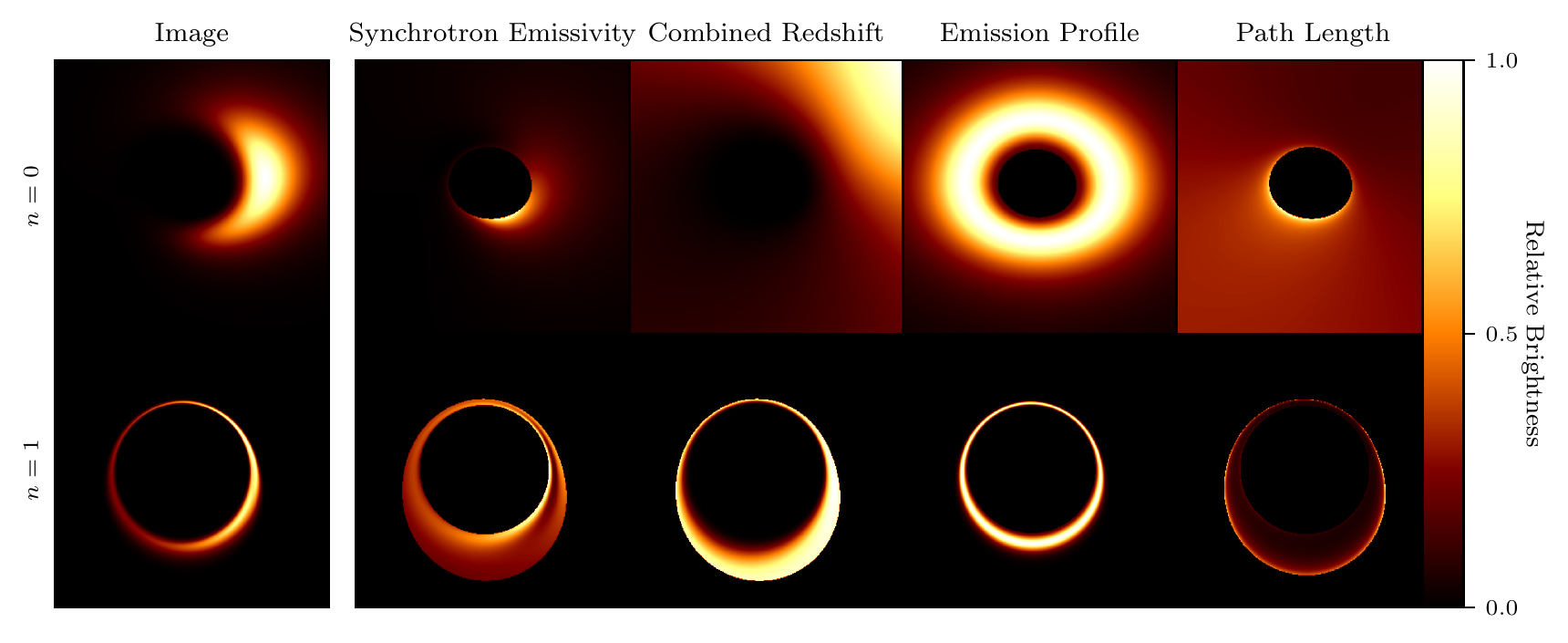}
    \caption{Example model images (first column) and their factorization into four underlying quantities: synchrotron emissivity (second column), combined Doppler and gravitational redshift (third column), emissivity profile (fourth column), and optically thin path length factor. \edit1{The relative brightness here and in all other figures is normalized in individual panels.} Underlying model parameters are $\theta_{\rm g}=3.8$ $\mu$as, $a_* = -0.5$, $\theta_{\rm o} = 45^\circ$, $\beta=0.5$, $\chi=-3\pi/4$, $\iota=\pi/3$. The emissivity profile has $R=4.5$ M, $p_1=p_2=3$. Top and bottom rows show quantities for the $n=0$ and $n=1$ images, with increasing $n$ corresponding to subsequent half-orbits of light around the black hole. }
    \label{fig:mult_decomp}
\end{figure*}

The model is diagrammed in \autoref{fig:pedagogy}. In detail, we assume an axisymmetric flow with constant speed and magnetic field magnitude producing only synchrotron emission for which the following parameters (measured in the ZAMO frame) are uniform across the equatorial plane: the fluid speed $\beta$ expressed as a fraction of the speed of light, position angle $\chi$ of the flow's velocity with respect to the radial unit vector, spherical polar angles $\eta$ (azimuthal) and $\iota$ (polar) specifying the orientation of the magnetic field, and spectral index $\alpha_\nu$. We follow \citetalias{Narayan_2021} and take $\eta=\chi+\pi$ (assuming magnetic flux freezing, the inner accretion flow drags field lines behind the velocity vector). The black hole angular gravitational radius $\theta_{\rm g} \equiv G M / Dc^2 $  (also known as the mass-to-distance ratio, expressed in $\mu$as), dimensionless spin $a_*$, and observing inclination $\theta_{\rm o}$ determine the geodesics that correspond to particular pixels in the image grid. As in \citetalias{Gelles_2021}, we take $\theta_{\rm o} < 90^\circ$ but allow a signed spin $a_*$ so that a negative spin is oriented away from the observer, and $\chi=-\pi/2$ will always be clockwise on the sky. 
Our convention differs from the EHT papers, in which positive/negative spin indicates a prograde/retrograde accretion flow; we will thus describe GRMHD simulations using our convention and always specify whether the accretion flow is prograde or retrograde when $a_*$ is non-zero.
All potentially fit parameters are summarized in \autoref{tab:params}.  Typically, we fix the angular field of view (FOV), the number of pixels \npix, and the maximum number $n_{\rm max}$ of lensed sub-images to include in the forward model.

We include one significant departure from the \citetalias{Narayan_2021} construction: we introduce an additional cross product spectral index $\alpha_\zeta$ which characterizes the dependence of the observed intensity on the quantity $\sin\zeta \propto |\vec{k} \times \vec{B}|$, where $\vec{k}$ is the spatial momentum of an emitted photon and $\vec{B}$ is the fluid frame magnetic field. The emitted intensity then goes as $I_\nu \propto \sin(\zeta)^{(1+\alpha_\zeta)}$. In \citetalias{Narayan_2021}, $\alpha_\zeta = \alpha_\nu$, and the choice of $\alpha_\nu$ is informed by the electron distribution function. The fiducial $\alpha_\nu=\alpha_\zeta=1$ used in \citetalias{Narayan_2021} was taken from GRMHD intuition; we allow them to vary independently to capture potentially complicated behavior when the source of interest is in the transition region between optical thinness and thickness. In general, we should expect $\alpha_\zeta$ to be close to $\alpha_\nu$ in fits to GRMHD.

We trace rays semi-analytically using the elliptic formalism given in appendix B of \citet{Gralla_2020_null}, which showed that inverting from screen coordinates to Boyer-Lindquist radius only requires evaluation of Jacobi trigonometric functions and elliptic integrals of the first kind on real arguments (see \citet{Rauch_1994} and \citet{Dexter_2009} for similar treatments).  Recovering $\phi$ generally requires evaluation of the Legendre incomplete elliptic integral of the third kind, which is significantly more computationally taxing. Thus, for simplicity and speed, we avoid computation of $\phi$ by only forward-modeling axisymmetric emissivity distributions. Asymmetry in sky images is thus produced solely by the inclination of the disk, causing variation in the synchrotron emission (due to changing angles between magnetic fields and geodesics), Doppler and gravitational redshift factors, and optical path length.

After tracing rays, the model has specified a synchrotron emissivity for all points in the black hole equatorial plane; we then multiply emissivities by an envelope function $\mathcal{J}(r,\phi)$ ($\mathcal{J}(r)$ when axisymmetry is assumed), evaluated over the ray traced Boyer-Lindquist radial and azimuthal coordinates $r$ and $\phi$ at the midplane. This emissivity profile is the sole explicit source of radial structure in the model specification, implicitly representing variation in the magnetic field strength $|\vec{B}|$, electron density $n_e$, and temperature $T_e$, as well as geometric factors such as disk thickness.

We allow the axisymmetric envelope $\mathcal{J}(r)$ to take any number of additional fitted parameters specifying variation of the envelope over radius, supplied at runtime; for the examples shown in this paper, we use a ring with the functional form of a double power law,
\begin{align}
    \label{eqn:profile}
    \mathcal{J}(r; R, p_1, p_2) &= \frac{(r/R)^{p_1}}{1+(r/R)^{p_1+p_2}},
\end{align}
where $R$ specifies the scale radius and $p_1$ and $p_2$ modulate the steepness of the power law in the interior and exterior of $R$, respectively, with larger values increasing steepness. At small radii, this function goes like $(r/R)^{p_1}$, while at large radii, it goes like $(r/R)^{-p_2}$.

The synchrotron emissivity, combined Doppler and gravitational redshift, emission profile, and path length factor jointly specify the image through a simple product, as shown in \autoref{fig:mult_decomp} for a steeply inclined ($\theta_{\rm o}=\pi/4$) model. We show image quantities decomposed for the $n=0$ and $n=1$ images, where $n$ indexes the number of half-orbits photons underwent before reaching the observer (see \citet{Johnson_2020} for a detailed treatment of photon rings). 

We observe that both the redshift and synchrotron emissivity contribute asymmetry at a similar dynamic range. Further, we see that the synchrotron emissivity increases towards the black hole.
This effect arises from the magnetic field geometry; with $\chi=-3\pi/4$ and $\iota=\pi/3$, the magnetic field is half vertical and half equatorial, and the equatorial component is half radial and half azimuthal. This combination means that photons emitted with momentum aligned with the equatorial plane (favoring near-horizon emission) will have the largest $\vec{k}\times\vec{B}$ due to being perpendicular to the vertical direction. 

We also note that the path length increases towards the image center in the $n=0$ image and away from the image center in the $n=1$ image. The former can be explained similarly to the emissivity, in that near-horizon emission that reaches the observer must be emitted with near-radial momentum, which has \edit1{a} large path length at the midplane. The latter is indicative of photons emitted nearly radially inward by material at very large radii.
The other quantities in \autoref{fig:mult_decomp} are more intuitive; the east-west asymmetry in the shape of the $n=1$ quantities is given by the non-zero black hole spin, while the brighter side of the redshift is set by the clockwise motion of the fluid on the sky (as given by a negative $\chi$).

After computing the emission properties, the resulting image is modified by two post-processing parameters which also may vary in fitting: the total Stokes $I$ flux density (hereafter $I_{\rm tot}$) and the position angle (PA) of the projected black hole spin axis on the sky (corresponding to a simple image rotation). 
Enforcing the total flux $I_{\rm tot}$ makes normalization of the emissivity function irrelevant.

Lensed sub-images require special care. Computing the radius at which a geodesic encounters the equatorial plane in general requires knowing the full geodesic as well as the Mino time $\tau$ at which the impact occurs, as we describe later in \autoref{subsec:rays}. We thus need to compute the emission radius for a given number of windings about the black hole $n$, which requires computing the Mino time $\tau_n$ of each successive midplane crossing, achievable through Equation 17 of \citetalias{Gelles_2021}. This equation will readily yield unphysical Mino times that exceed the total Mino time $\tau_{\rm tot}$ of the entire geodesic. Thus, we compute $\tau_{\rm tot}$ once per model evaluation for the full image, and only compute lensed image regions when $\tau_n < \tau_{\rm tot}$. The result is a shrinking region that rapidly approaches the critical curve with increasing $n$, as geodesics with very large Mino time correspond to rays that wrap around the black hole many times.

\subsection{Ray tracing sub-regimes}
\label{subsec:rays}

As discussed in detail in \citet{Gralla_2020_null}, null geodesics in a Kerr spacetime can be grouped into four cases corresponding to which of the four roots ($r_1$, $r_2$, $r_3$ and $r_4$) of the quartic radial potential are real or imaginary. As treated explicitly in appendix A of \citet{Gralla_2020_lensing}, the rays relevant to tracing our forward model are all case (1), (2) or (3), as restricting emission to the equatorial plane removes vortical geodesics and those that do not encounter the midplane. In cases (1) and (2), all roots are real, while in case (3), $r_3$ and $r_4$ are complex and obey $r_3 = r_4^*$.

\citetalias{Gelles_2021} uses the unified inversion formula for radius given at the end of Appendix B of \citet{Gralla_2020_null}; this formula works over all cases by allowing complex arguments in the Jacobi elliptic functions. We restrict to real arguments for computational speed and simplicity, and thus use two different inversion schemes for cases (1) and (2) and case (3), corresponding to the case-specific formulas given in \citet{Gralla_2020_null} as a function of  $\tau$. Denoting cases with superscripts, we use
\begin{align}
    \label{eqn:inversions}
    r^{(1,2)}(\tau) &= \frac{r_4(r_3-r_1) - r_3(r_4-r_1){\rm sn}^2\left(X_2(\tau)|k_2\right)}{(r_3-r_1)-(r_4-r_1){\rm sn}^2\left(X_2(\tau)|k_2\right)},\\
    r^{(3)}(\tau) &= \frac{(B r_2 - A r_1) + (B r_2 + A r_1){\rm cn}(X_3\left(\tau)|k_3\right)}{(B-A)+(B+A){\rm cn}(X_3\left(\tau)|k_3\right)},
\end{align}
where $A$, $B$, $k_2$, and $k_3$ are simple algebraic functions of quartic roots and $X_2$ and $X_3$ are functions of the Mino time and screen coordinates; all are real, and can be found in \citet{Gralla_2020_null}. Our implementation of the case-based ray tracing also borrows heavily from the implementation in \texttt{kgeo} \citep{Chael_kgeo_2022}, recently used in \citet{Levis_tomography}, a machine learning-based approach to the inference of emission geometry in a fixed Schwarzschild spacetime.

When tracing rays, our code permits tracing different sub-images $n$ at differing resolution; for all fits to VLBI data, we use \edit1{a} 80 $\mu$as field of view with square pixels of side length 2 $\mu$as  in the $n=0$ image and 2/3 $\mu$as pixels in the $n=1$ image, unless otherwise noted. We avoid high-resolution computation of geodesics which do not impact the midplane at a given $n>0$ by first computing the sub-image Mino time at the resolution of the previous ($n-1$) sub-image, yielding a coarse mask of rays which are supported at sub-image $n$; the full ray-tracing calculation is only carried out for pixels in this mask or within 1 pixel of this mask, re-sampled at the higher resolution. This approach guarantees that at the higher resolution, no rays which contribute flux are missed despite the hierarchical resolution mask.

\subsection{Expectations and Limitations}

\begin{figure*}
    \centering
    \includegraphics[width=\textwidth]{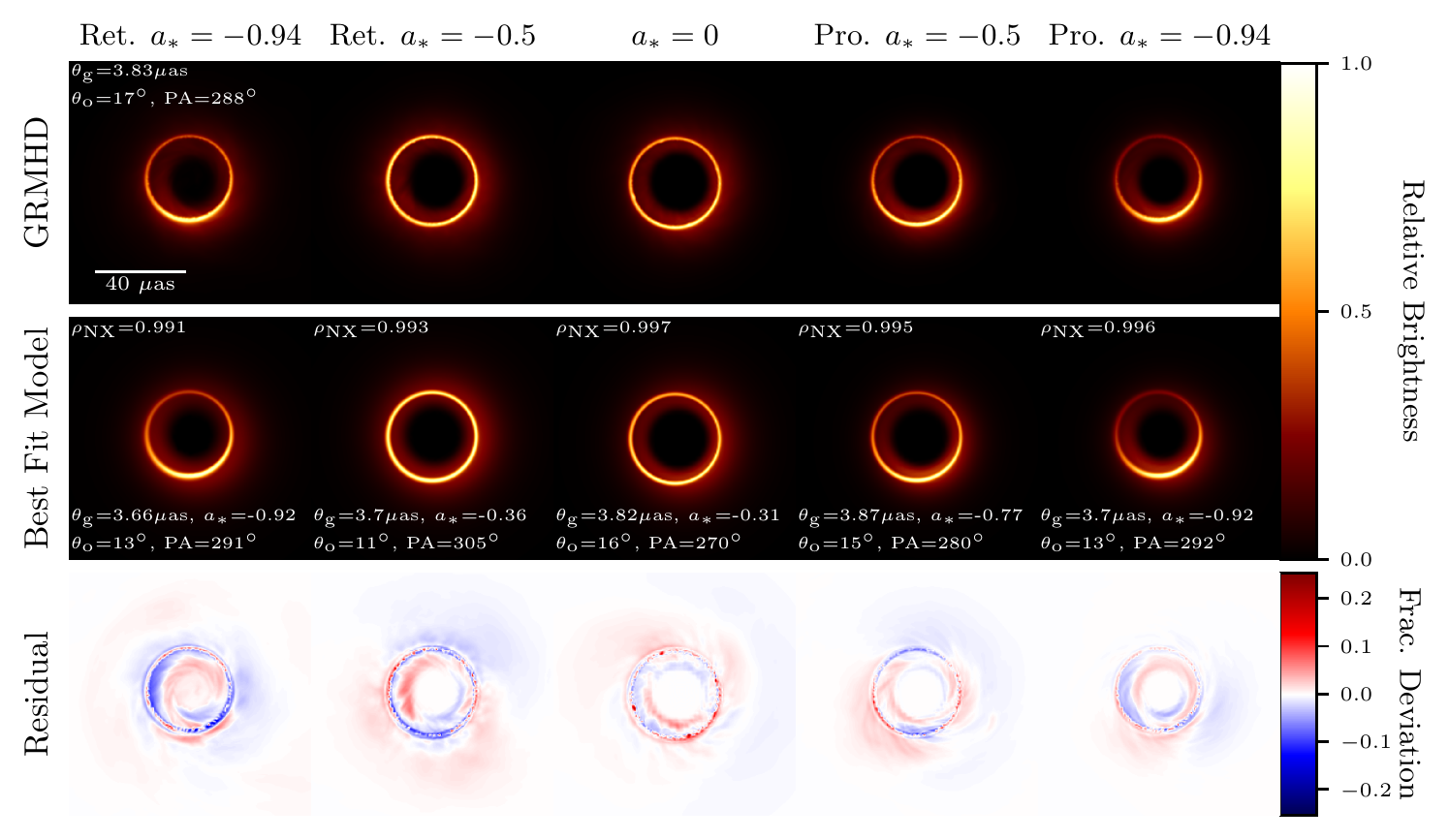}
    \caption{Comparison of time-averaged images of magnetically arrested GRMHD simulations of various spins with $R_{\rm high} = 20$ and the corresponding best-fit model image with $n_{\rm max}=1$ from a maximization of the normalized cross correlation $\rho_{\rm NX}$. The bottom row shows residuals, where blue indicates that the model fit exceeds the GRMHD image, and red indicates the opposite; residuals are expressed as a fraction of the peak Stokes $I$ flux of the GRMHD image. Model images are annotated at the top left with the best-fit $\rho_{\rm NX}$ and at the bottom left with the inferred black hole and viewing parameters.}
    \label{fig:mr20}
\end{figure*}

\begin{figure*}
    \centering
    \includegraphics[width=\textwidth]{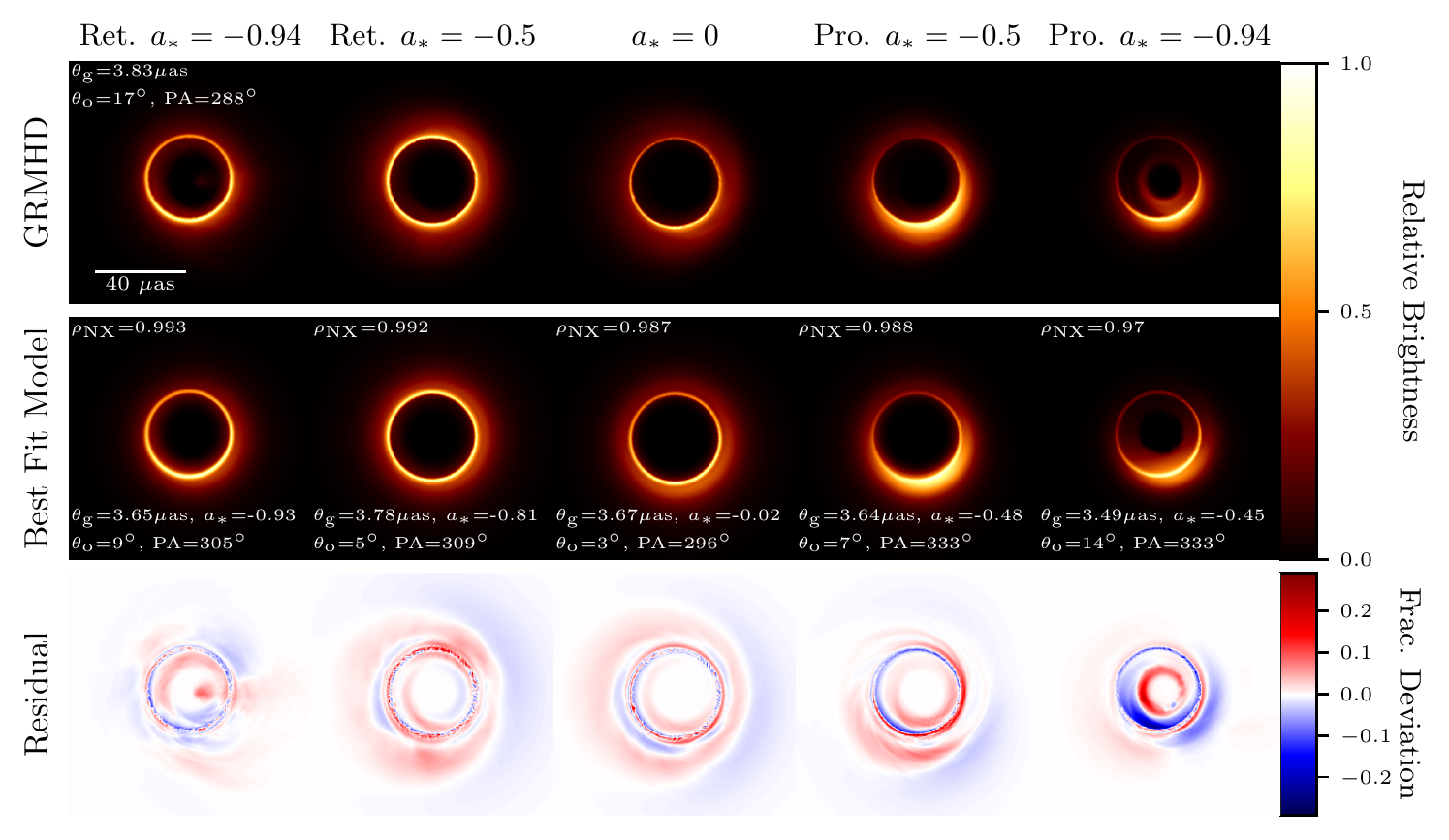}
    \caption{Same as \autoref{fig:mr20} but for SANE simulations with $R_{\rm high}=160$, for which the majority of emission is far above the Kerr midplane, violating the assumptions of our toy model. }
    \label{fig:sr160}
\end{figure*}

Our primary targets of comparison will be simulations in the \citetalias{PaperV} library, which contains GRMHD flows in both the Magnetically Arrested Disk (MAD) and Standard and Normal Evolution (SANE) states. MAD flows typically have stronger, more ordered magnetic fields (sufficiently strong to balance the inward ram pressure of the gas) and accrete more efficiently than SANEs \citep{Ichimaru1977,Igumenschchev_2003,Narayan2003, Narayan_2012, Yuan_and_Narayan_2014}. Most saliently, MADs and SANEs also differ in emission geometry even when all other input parameters are the same, as discussed in \citetalias{PaperV}. 
In addition to the two accretion states, the library also contains models of five different dimensionless spins vectors, defined relative to the angular momentum of the large-scale accretion disk. In our notation, in which a negative value of $a_*$ refers to a spin vector oriented away from the observer, the library spin values can be most unambiguously described as retrograde $a_*=-0.94$, retrograde $a_*=-0.5$, $a_*=0$, prograde $a_*=-0.5$, and prograde $a_*=-0.94$. 
The library contains six values of the electron heating parameter $R_{\rm high}$, which post-processes the electron temperature relative to the ion temperature produced by GRMHD \citep{Mosci_2016}. Higher values of $R_{\rm high}$ typically elevate the dominant emission region out of the equatorial plane (particularly for SANEs) and reduce polarization. Lastly, due to the focus on \m{}, the library contains only models with a viewing inclination of $17^\circ$.

We use time-averaged images of the full library ray traced at high resolution (480 pixels across 160 $\mu$as on each edge) and updated mass-to-distance ratio in \citet{Palumbo_2022}. The GRMHD simulations were done with \texttt{iharm3D} \citep{Gammie_HARM_2003, IHARM3d_prather}; ray tracing was performed using \texttt{ipole} \citep{IPOLE_2018}. Additional details on the image generation process may be found in \citet{Wong_2022}. In all cases, we rotate images so that the approaching jet (and negative spin axis position angle) is oriented $288^\circ$ degrees east of north.

Before fitting models to synthetic VLBI data, it is constructive to examine the capability of the model to resemble GRMHD simulations despite the many simplifying assumptions in our construction. To do so, we conduct fits in the image domain using \texttt{scipy}'s dual annealing to maximize the normalized cross correlation $\rho_{\rm NX}$ \citep{PaperIV}:
\begin{align}
    \rho_{\rm NX} \equiv \frac{1}{N}\sum_i \frac{(X_i-\langle X\rangle)(Y_i-\langle Y\rangle)}{\sigma_X \sigma_Y}.
\end{align}
Here, angle brackets denote a pixel-wise average of the Stokes $I$ flux and $\sigma_X$ and $\sigma_Y$ denote the standard deviation of the pixel values of the Stokes $I$ flux in each image. The summation indexes pixels with $i$. In order to examine large scale residuals in the intensity, we use a larger FOV and $N_{\rm pix}$ here than in later fits to VLBI data, with a FOV of 160 $\mu$as across 240 pixels.

\autoref{fig:mr20} shows image-domain fits of the model to time-averaged images of MAD GRMHD simulations with $R_{\rm high} = 20$ of varying spin. \autoref{fig:sr160} shows similar fits to SANE simulations with $R_{\rm high}=160$. The low-$R_{\rm high}$ MADs can be taken as an optimistic case for our model due to their more ordered fields and more planar emission geometry, while the high-$R_{\rm high}$ SANEs are a priori more challenging due to their brighter off-plane emission. These fits then serve as a proxy for perfect observations with complete Fourier sampling in an interferometric context and thus set an upper bound on the model's capability to mimic an image. We find that in the MAD fits, $\rho_{\rm NX}$ always exceeds 99\%, indicating nearly indistinguishable best fit images despite occasionally large differences in the inferred black hole parameters, which are annotated in the figure. Meanwhile, though the SANEs have slightly lower typical $\rho_{\rm NX}$ (particularly the prograde high spin SANE), the error in inferred parameters is much larger, most noticeably manifesting in an upwardly biased spin axis position angle across all spins.

We conclude that the toy model has no trouble resembling the quiescent structures of GRMHD simulations; model misspecification will then manifest as images which resemble the source at the resolution of the observing instrument with potentially biased inferred parameters. For the results in this paper (which fit only total intensity data), we then expect the following limitations:

\begin{enumerate}
    \item The model assumes an axisymmetric flow. This limitation will create untrustworthy results when fitting observations of non-axisymmetric flows, such as snapshots of GRMHD, while time-averaged GRMHD are nearly axisymmetric (though long-duration streamlines are still present, as can be seen in the residual plots in \autoref{fig:mr20} and \autoref{fig:sr160}).
    \item The model assumes all emission emerges from the Boyer-Lindquist midplane. In model fits to GRMHD simulations of \m{} containing bright emission from jet regions (such as SANEs with large $R_{\rm high}$), the effects of all model parameters on the observed angular radius of the ring will be biased.
    \item The model assumes that magnetic field and fluid velocity magnitude and orientation are both uniform in radius (in addition to axisymmetric), which will fail to reproduce structures emerging from radially extended emission geometries.
\end{enumerate}

These limitations generally arise from choices made to limit the number of model parameters and simplify model evaluation. The underlying assumptions can thus be relaxed in future work. We move forward with a very simple model so that we may characterize the necessity of a more complicated specification in light of fits to synthetic VLBI data from time-averaged GRMHD simulations.

\section{Example Model Fits to VLBI Data}
\label{sec:fits}

To assess the capabilities of the model, we perform two interferometric model fitting tests. First, we fit the model to itself using VLBI data, assessing fundamental degeneracies of the black hole lensing problem that vary as sub-images are added to the model. Then, we fit the model to GRMHD with $n_{\rm max}=1$ using synthetic VLBI data to assess parameter biases that arise from model misspecification. When fitting to data in both cases, we use log closure amplitudes and closure phases (and their associated thermal noise) from the interferometric sampling of the EHT's lower band viewing \m{} on April 11, 2017 to forward-model synthetic data \citepalias{PaperII}.

The parameter values and prior ranges used in each case are given in \autoref{tab:params}, each of which is tuned for the \m{} observational case. In general, we take wide priors aimed at capturing the ranges typically seen in GRMHD or in observations of \m{}. For $\theta_{\rm g}$, we allow a wide enough prior to comfortably encompass the $2\sigma$ excursions from stellar and gas dynamical estimates of the \m{} mass (see Table 9 of \citetalias{PaperVI}) because this prior width is appropriate for application to EHT data in the absence of other EHT-inferred parameters \citep{Gebhardt_2011, Walsh_2013}. For $a_*$, PA, and $\chi$, we restrict the prior volume again based on observations: the large scale jet indicates that the position angle of the inner accretion flow is oriented west, so we allow a $90^\circ$ window around the fiducial value taken by the EHT ($288^\circ$, see \citetalias{PaperV} and \citet{Walker_2018}). The brightness asymmetry in images requires clockwise motion ($\chi$ between $-\pi$ and $0$) which we assume to be aligned with the spin axis in the emission region (meaning the spin must be oriented away from the observer, that is, $-1<a_*<0$).
For all other parameters, we generally prescribe wide priors based on possible values in GRMHD simulations reported in \citetalias{PaperV} and \citetalias{PaperVIII} as well as \citet{Ricarte_2020}. All priors are uniform across their domain.

For all fits to interferometric data, we use \texttt{dynesty} \citep{Speagle_2020} to perform dynamic nested sampling of the posterior. \texttt{dynesty} is ideal for our model because it is well suited to low dimensional but highly degenerate and potentially multimodal posteriors, which are expected in the naturally degenerate black hole emissivity modeling problem. Throughout, we use $2500$ live points, at most 10 batches of dynamical nested sampling with 500 additional live points per batch (though with default stopping criteria, almost all fits conclude after the static nested sampling step), and a terminating change in the log of the evidence of $0.01$, ensuring a well-explored and well-resolved posterior volume. 


\begin{figure}
    \centering
    \includegraphics[width=0.47\textwidth]{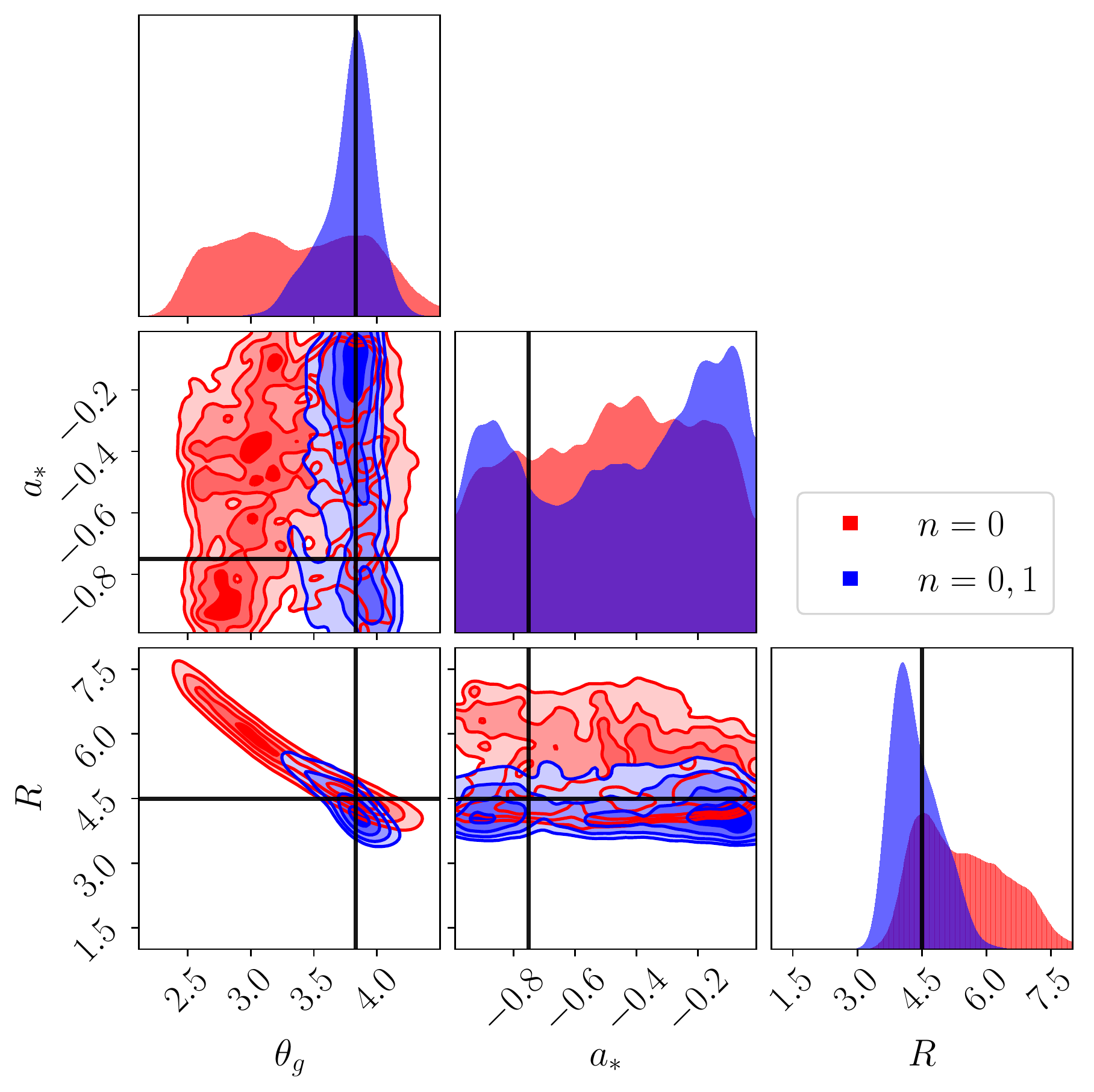}
    \caption{Partial triangle plots of posteriors on parameters of interest in a fit of the model to itself using closure quantities from 2017 EHT coverage of \m{}. In red, a model containing only the $n=0$ image is fit to itself; in blue, a model containing both the $n=0$ and $n=1$ image is fit to itself. True values are shown in black. 
    The underlying true model has $\theta_{\rm g}=3.83$ $\mu$as, $a_*=-0.75$, $\theta_{\rm o}=17^\circ$, ${\rm PA}=288^\circ$, $\beta=0.5$, $\chi=-\pi/2$, $\iota=\pi/3$, $\alpha_\nu=1$, $\alpha_\zeta=1$, $R=4.5$, $p_1=5$, and $p_2=5$, all of which are fit. Full posteriors are in \autoref{sec:fulln0n1}. For this coverage, addition of the $n=1$ image softens the degeneracy between black hole mass and emission radius in the fitted results but is still insufficient to constrain spin. Here and in all other triangle plots, contours show $0.5 \sigma$, $1 \sigma$, $1.5 \sigma$, and $2 \sigma$ levels.
    }
    \label{fig:n0_n1_selfcomp}
\end{figure}

\begin{figure}
    \centering
    \includegraphics[width=0.48\textwidth]{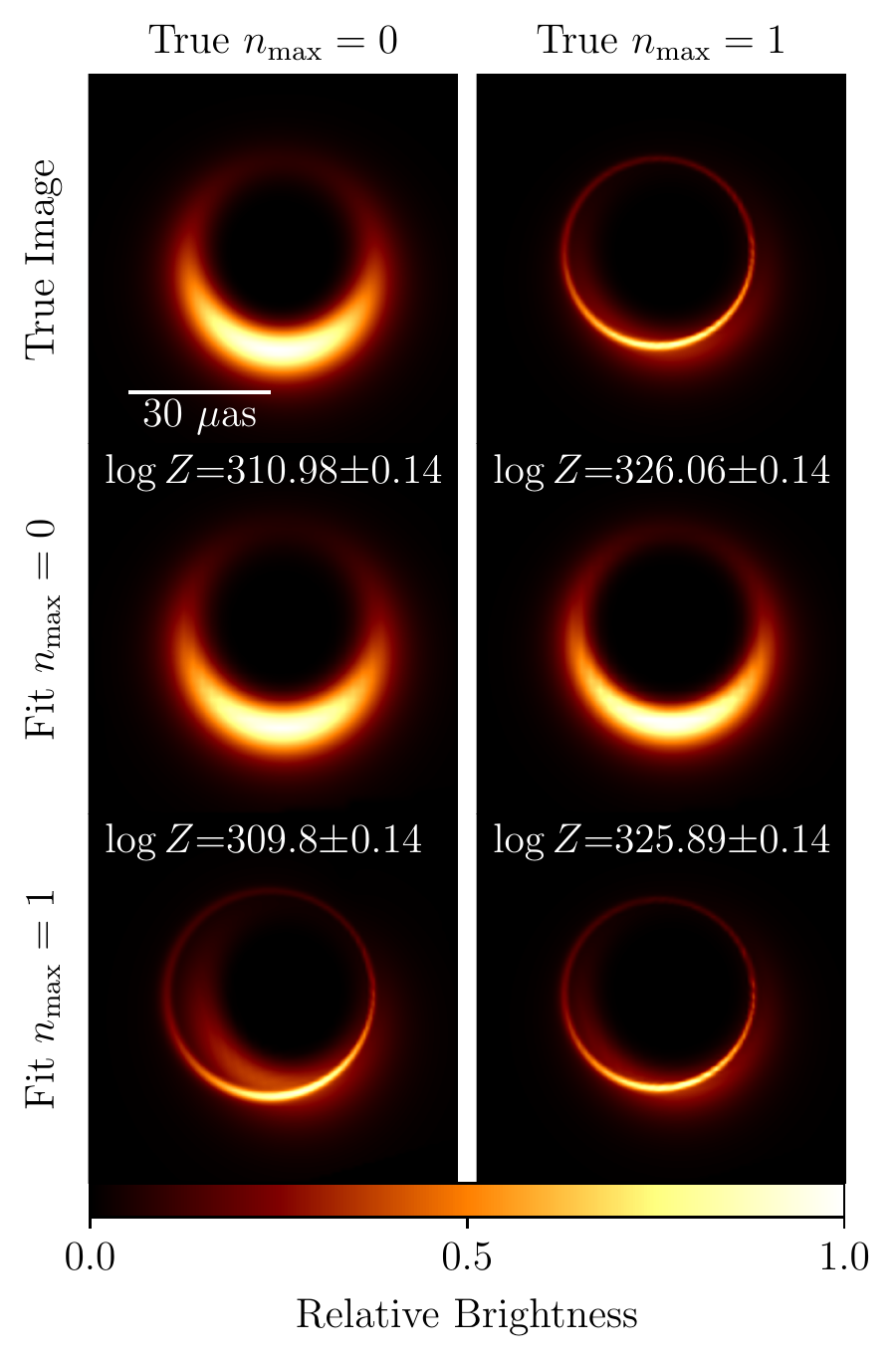}
    \caption{Posterior mean images of the toy model fit to itself with a matched and mismatched number of fitted sub-images annotated with the the log posterior evidence $\log Z$ estimated by \texttt{dynesty}, along with its estimated error. The underlying model and data are the same as in \autoref{fig:n0_n1_selfcomp}. Lower  $\log Z$ suggests a relatively disfavored model, though in the case of the fit to the $n_{\rm max}=1$ model, the incorrect specification is not favored statistically significantly.}
    \label{fig:n0_n1_grid}
\end{figure}

\subsection{Fitting the Model to Itself}

As higher $n$ sub-images approach the critical curve, we expect sensitivity to the absolute scale $\theta_{\rm g}$ of the accretion system that is inaccessible from the $n=0$ image alone. Further, the shape of the critical curve is a sensitive probe of spin \citep{Bardeen_1973, Takahashi_2004, Johannsen_2010}. We thus expect that model fits with the $n=1$ sub-image should show sharper constraints on $\theta_{\rm g}$, $a_*$, and all emission geometry parameters if the data are sensitive to the $n=1$ image structure. To assess the value of this model feature, we perform fits of the model to itself with identical input parameters and prior volumes, but with each of $n_{\rm max}=0$ and $n_{\rm max}=1$.

\autoref{fig:n0_n1_selfcomp} shows partial posteriors obtained from this self-fit. The fit with $n_{\rm max}=0$ is overlaid in red with the $n_{\rm max}=1$ posterior in blue; we observe that the addition of the $n=1$ sub-image better constrains $\theta_{\rm g}$ and $R$ by providing an absolute angular scale for the system, while the spin is left essentially unconstrained. 
The marginal sensitivity to the $n=1$ sub-image in EHT 2017 data and the wide range of variation of other parameters (see \autoref{fig:n0_n1_selfcomp_all} in \autoref{sec:fulln0n1}) likely limit the ability to constrain spin.

Self-fits offer us another opportunity. The most simple type of model misspecification we may assess is the fitting of a model with $n_{\rm max} = 1$ to one with $n_{\rm max}=0$, or the opposite. By changing $n_{\rm max}$ to 0 or 1 and fitting to a mismatched model, we may assess from data comparison metrics whether a particular model is preferred. Conveniently, changing $n_{\rm max}$ does not alter the prior volume or number of parameters in the fit, as the additional sharp ring is a natural consequence of the spacetime and not a new model component. Thus, a simple posterior model evidence comparison can be used to judge which of two values of $n_{\rm max}$ is preferred when fitting a single dataset.


\begin{figure*}
    \centering
    \includegraphics[width=\textwidth]{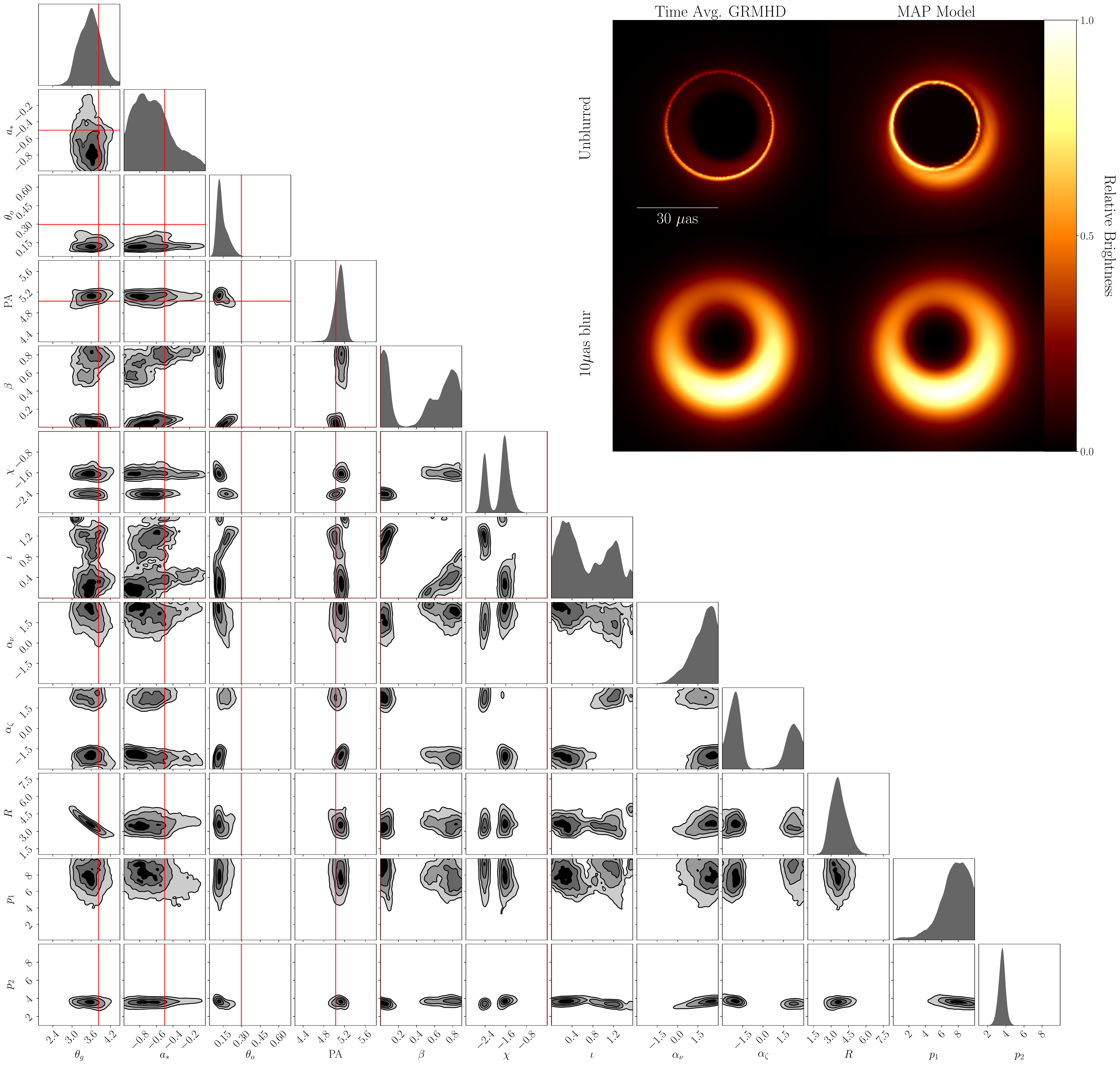}
    \caption{Bottom left: posteriors on all fit parameters for 2017 EHT coverage of a MAD with prograde $a_*=-0.5$ and $R_{\rm high}=20$. Red lines indicate true values for the extracted GRMHD simulation and ray tracing parameters. Top right: time-averaged image of the GRMHD simulation and maximum likelihood sample from the posterior, shown with and without a 10 $\mu$as blur.}
    \label{fig:GRMHD_corner}
\end{figure*}

\begin{figure*}
    \centering
    \includegraphics[width=0.98\textwidth]{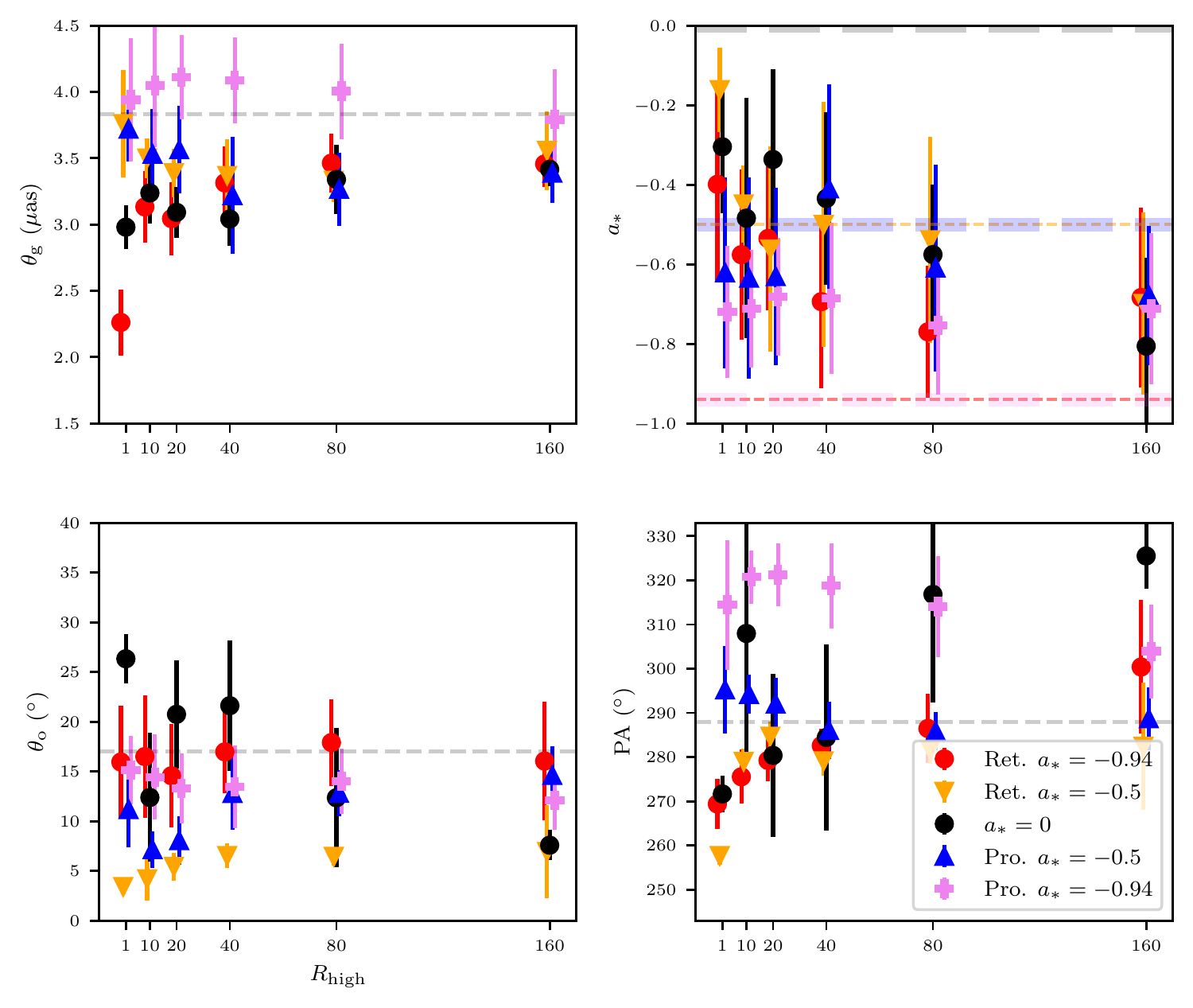}
    \caption{Summary of model fits to synthetic data from time-averaged MAD GRMHD simulations of the \m{} accretion flow with $1\sigma$ error bars. True values are shown with dashed lines, while color indicates spin. Small horizontal offsets near a single $R_{\rm high}$ are for visual clarity. \edit1{Vertical ranges correspond precisely to prior ranges.} With the exception of prograde high spin models, most inferred masses and viewing inclinations are lower than the true value. Notably, non-spinning models rarely show a strong posterior preference for low spins, though spin is almost always unconstrained.}
    \label{fig:MADs}
\end{figure*}

\begin{figure*}
    \centering
    \includegraphics[width=0.98\textwidth]{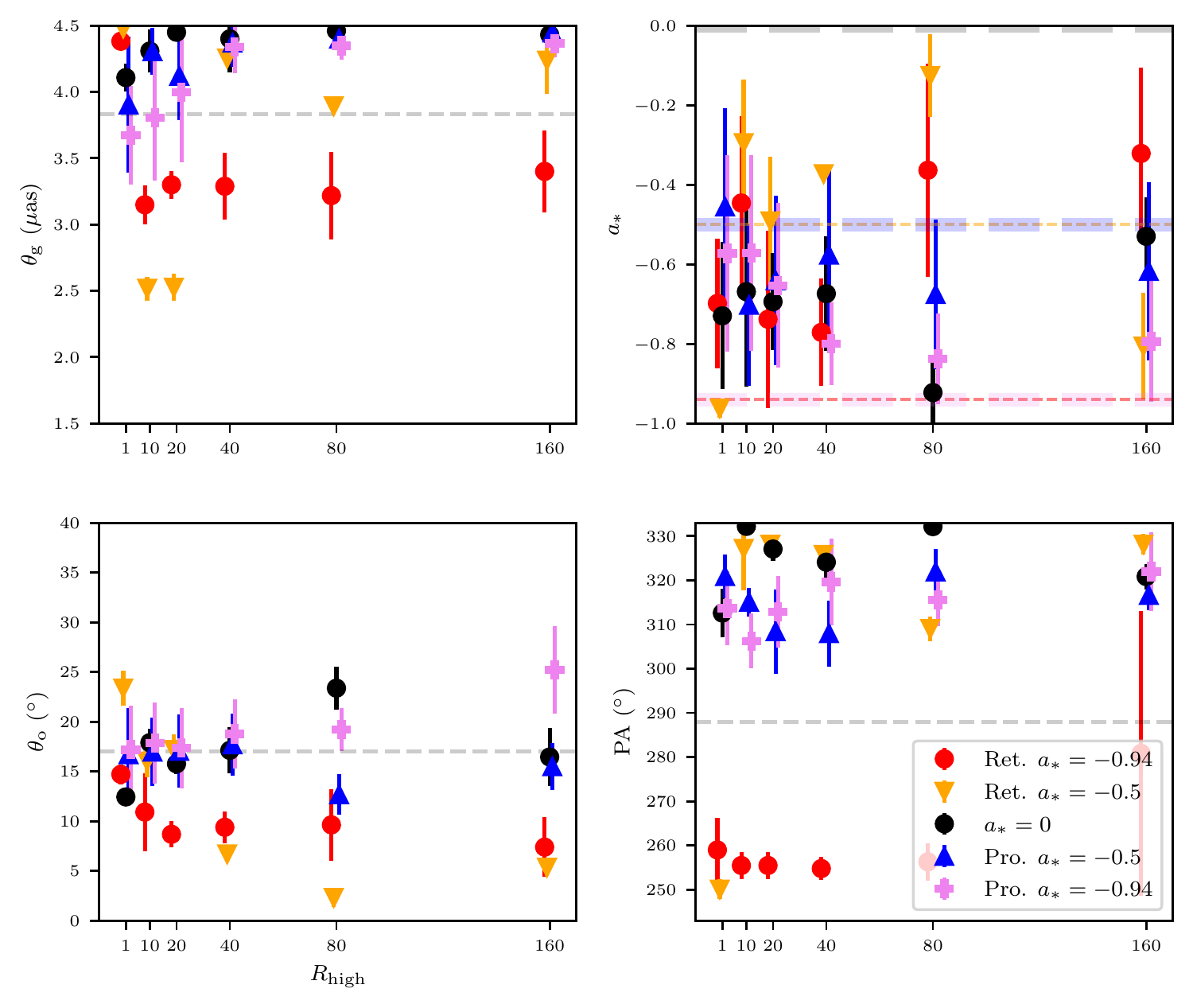}
    \caption{Same as \autoref{fig:MADs} but for time-averaged SANE GRMHD simulations. Due to the worse model misspecification for SANEs, many fits show inaccurate yet certain measurements of model parameters. The tendency for results to worsen with increasing $R_{\rm high}$ indicates that the cause is emission moving off of the equatorial plane at higher $R_{\rm high}$, violating the main model assumption of equatorial emission.}
    \label{fig:SANEs}
\end{figure*}

\autoref{fig:n0_n1_grid} shows the posterior mean models and log model evidence fit under four possible model (mis)specification scenarios, in which correct and incorrect numbers of sub-images are included when fitting. We see that adding the $n=1$ image is weakly disfavored when the true model does not contain it, while failing to include the $n=1$ image when it ought to be present is very slightly favored. We may estimate the relative probability of the two models from the difference in the log model evidence $\log Z$; in both cases, $\exp(-\Delta \log Z)>0.05$. Thus, the correct model is not preferred with statistical significance in either case, showing that the EHT 2017 coverage and sensitivity do not distinguish the presence or absence of the $n=1$ ring; however, when the ring is present in the underlying image, the posterior tightens around the true parameters and replicates the correct ring size. This result is likely sensitive to the amount of flux in the $n=1$ ring, which itself depends delicately on the physical parameters of the flow. 
These fits suggest that the 230 GHz EHT is marginally sensitive to the photon ring on its longest baselines, observing comparable visibility amplitude in the the $n=0$ and $n=1$ images (see discussion in \citet{Johnson_2020}). 

\subsection{Fitting the Model to time-averaged GRMHD Images}

\label{subsec:grmhd_fitting}

In all model fits to GRMHD simulations, we add a modest $1\%$ fractional systematic noise on complex visibilities, which propagates into log closure amplitudes and closure phases. Equivalently, thermal noise uncertainties $\sigma_{\rm th}$ on individual visibility amplitudes are added in quadrature with a new noise term $\sigma_f$ determined by the fractional systematic error $f$:
\begin{align}
    \sigma_{f,i} &\equiv f |V_i|,
\end{align}
where the index $i$ indicates distinct measurements and $|V_i|$ is an individual measured visibility amplitude. This noise budget prescription is consistent with those used in \citetalias{PaperIV} and \citetalias{PaperVI}. The systematic noise component is not included in the debiasing of log closure amplitudes (see \citet{TMS} for details on debiasing of the amplitudes of complex data).

As before, we generally fit a FOV of 80 $\mu$as with 120 pixels along an edge. However, SANE models with $R_{\rm high} =1 $ have very large angular extent; for these models, we double the field of view to 160 $\mu$as and the number of pixels to 240 along an axis. In general, informed choices of the field of view are realistic even in the case of real data using pre-imaging arguments from the observed visibility amplitude, as in the pre-imaging analysis in \citetalias{PaperIV}.

First, we examine the case of a time-averaged GRMHD simulation where we expect the model to be a decent approximation of the quiescent flow; that is, a MAD model with low $R_{\rm high}$. As discussed in \citet{Chael_2021} and \citetalias{PaperV}, emission from MADs tends to emerge primarily from near the midplane, particularly at low values of $R_{\rm high}$. Further, since MADs tend to have stronger and more ordered magnetic fields than SANEs, we expect time averaging to preserve large scale features of the synchrotron emissivity, whereas in SANEs with unstructured magnetic field, the time-averaged image may correspond to vastly different underlying photon momenta-magnetic field cross products $\vec{k}\times \vec{B}$ than in any individual snapshot.

The posterior and maximum likelihood sample for the fit of a MAD simulation with prograde $a_*=-0.5$ and $R_{\rm high}=20$ is shown in \autoref{fig:GRMHD_corner}. 
For the known parameters of the simulation ($\theta_{\rm g}$, $a_*$, $\theta_{\rm o}$, and PA), we find multi-modal posteriors that are generally consistent or close to consistent with the truth. The fit has two modes with similar $\theta_{\rm g}$ posteriors. $a_*$ is nearly unconstrained, while $\theta_{\rm o}$ appears biased downward. The bimodal nature emerges most in $\beta$, $\chi$, and $\alpha_\zeta$; the lower $\beta$ mode with positive $\alpha_\zeta$ is likely closer to reality in the GRMHD. Recovered values for other parameters are either unconstrained or physically reasonable, with a clockwise rotating fluid $\chi$, and an emission radius consistent with the ranges documented in \citetalias{PaperV}. Most strikingly, we see that at 10 $\mu$as resolution (approximately half of the nominal resolution of the 2017 \m{} coverage of the EHT), the model is indistinguishable from the input image.

From the wide $\theta_{\rm g}$ posterior and unconstrained $a_*$, we may conclude that the photon ring is not dominating parameter inference. Moreover, the preference for a large $p_1$ (corresponding to a sharp inner edge in the accretion flow) suggests that the limited sampling of the 2017 EHT is insufficient to distinguish similar sources of high spatial frequency structure.

Next, we examine the success of parameter inference across the entire library. \autoref{fig:MADs} shows results for all MAD simulations, while \autoref{fig:SANEs} shows SANE results. Across MADs and SANEs, we observe several general trends:

\begin{enumerate}
    \item The model does not produce a narrow posterior around the true value of $\theta_{\rm g}$, suggesting that between the model misspecification and the 1\% fractional systematic error, no absolute angular sensitivity is achieved through sensitivity to the photon ring.
    \item Black hole spin is generally not constrained, as the full range of possible values is almost always permitted at the $2\sigma$ level.
    \item The viewing inclination is often underestimated (favoring more face-on viewing than truth).
    \item Retrograde GRMHD simulations tend to produce slightly worse accuracy in fit parameters.
\end{enumerate}

Between the MAD and SANE survey, we observe that MAD fits tend to be consistent at the $2\sigma$ level with the true values across nearly all models, with retrograde $a_*=-0.5$ models as the main exception. This result is in stark contrast to the SANE fits, where the majority of models (particularly those with $R_{\rm high} > 40$) are inconsistent with the true mass and position angle. For all SANEs except those with retrograde $a_*=-0.94$, the PA is overestimated in similar magnitude to the favored values in the earlier $\rho_{\rm NX}$ maximization. The primary difference between MADs and SANEs in time-averaged images is the emission geometry, so we attribute the difference in the model fits to the brighter off-equatorial structure in SANEs (see Figures 2, 3, and 4 of \citetalias{PaperV}).

Many fits exhibit certainty in recovered parameters even when the recovered parameters are grossly inaccurate. This result is suggestive of effects that fundamentally separate the toy model from (even time-averaged) GRMHD, manifesting as an insufficiency of the 1\% fractional systematic error to absorb all model misspecification. Taken as a whole, the GRMHD library fits provide a realistic collective test of the model's accuracy. Even in the generous case where a time-averaged image of the \m{} flow is available, our results nonetheless demonstrate that additional model complexity is required to capture the true underlying spacetime parameters alongside the persistent emission structure of emission.

\section{Discussion}
\label{sec:discussion}

The strongly curved spacetime near a black hole results in subtle and intricate effects on images of the surrounding plasma, which can be challenging to decouple from the signatures of a complex accretion environment. 
However, the remarkable simplicity of the Kerr spacetime---described fully by the black hole's mass and spin---allows for efficient modeling and direct inference of the black hole properties. 

We have developed an accretion flow model fitting framework, expanding work by \citetalias{Narayan_2021} and \citetalias{Gelles_2021}. Our approach uses an exact, semi-analytic description of null geodesics in the Kerr spacetime together with a simple geometric representation of the accretion flow that is informed by GRMHD simulations. By assuming optically thin, equatorial emission, we obtain a semi-analytic forward model that connects the black hole and accretion properties to their images. We showed that the model is successful in replicating images of time-averaged GRMHD simulations.
We used this scheme to forward model interferometric observations of black hole accretion flows, enabling rapid posterior estimation of both black hole and accretion flow parameters simultaneously. Example fits of the model to itself successfully captured the fundamental degeneracies between the spacetime and emission profile in the problem. Model fits that contained the $n=1$ photon ring showed narrower posteriors on black hole parameters even with the limited sampling of 2017 EHT data, though the data are insufficient to demonstrate the presence of the $n=1$ ring. 
Model fits to time-averaged GRMHD simulations indicated that the model is significantly better suited to fitting MAD models than SANES. In MAD models, most inferred parameters were consistent with true values at the $2\sigma$ level, while in SANEs, particularly with large $R_{\rm high}$, inferred values of mass and position angle trended high, occasionally saturating prior bounds with highly certain, highly inaccurate results.

Our axisymmetric synchrotron emitting plane model is a highly simplified prescription 
for the accretion flow in sources such as \m and \s, representing a projection of the accretion flow and outflow region onto the equatorial plane. As such, we see many avenues for additional, feasible complexity. For example, as shown in \citetalias{PaperV}, snapshots of GRMHD simulations are generally not axisymmetric, and non-planar emission geometries are also perfectly capable of producing asymmetric rings. Though sufficient averaging may permit robust fitting of axisymmetric models to real data, an expanded model could account for a number of features of realistic flows while still avoiding numerical radiative transfer. At the very least, we expect that future improvements to our formalism will include:
\begin{enumerate}
    \item emitting material above and below the midplane, efficiently modeled via two-dimensional sheets, perhaps as cones in Boyer-Lindquist coordinates;
    \item radial structure in fluid velocity and magnetic field;
    \item non-zero absorptivity;
    \item a vertical velocity component;
    \item non-axisymmetric structure, requiring evaluation of additional eliptic functions;
    \item fitting polarization as well as total intensity, which provides additional constraints on the magnetic field and spacetime \citep[see, e.g.,][]{Himwich_2020,Narayan_2021, Palumbo_2022}.
\end{enumerate}

Our approach, and other emissivity modeling approaches, benefit from the fact that sharp photon ring features are produced as a natural consequence of the spacetime, meaning that the dimensionality of the spacetime model is innately resolution invariant. The model fitting strategy employed by \texttt{KerrBAM} would thus not need to change when supplied with data including heterogeneous spatial frequency coverage, such as that which might come from an EHT array augmented with a space dish.

Realistic application of the model to \m{} data is limited in part by time-evolving non-axisymmetric structure in snapshots of the accretion flow. This challenge may be addressed through non-axisymmetric (and potentially time-evolving) emission envelopes $\mathcal{J}(r,\phi)$, or through variability noise budgets such as those in \citet{Broderick_2022} and motivated in \citet{Georgiev_2022}. \edit1{Most promisingly, recent advances in frequency phase transfer \citep[see, e.g.][]{RD_review} may enable coherent averaging of EHT visibilities over a series of observations, enabling direct fitting of the average flow. However, even in fits of the model to time-averaged images, the black hole spin is unconstrained with EHT 2017 data. Due to the insensitivity of the model to the many dynamical signatures of spin, traction on spin with \texttt{KerrBAM} is fundamentally limited by degeneracies between mass, spin, and emission location and their effect on the shape and size of the photon ring. At near face-on viewing, it is likely that even significantly more complete VLBI coverage will not sharply constrain spin in the time-averaged Stokes $I$ image (as evidenced by the image-domain fits to SANEs). However, additional constraints from polarimetry of the $n=0$ and $n=1$ images may break these degeneracies by tracking the frame dragging of magnetic fields near the horizon \citep{Palumbo_2022}.}

We have focused on model fits appropriate for \m{}, and so have prescribed prior ranges for the \m{} problem; the exquisitely known mass of \s{} \citep{Gravity_2019} would cleave through mass degeneracies, while the much less clear spin, viewing inclination, and spin axis position angle of \s{} would present a much wider and more degenerate posterior space on angular parameters, including those in the flow such as $\chi$ and $\iota$. Moreover, the observational prospects for \s{} are much more likely to produce a statistically sound ``average'' accretion state due to the number of dynamical times we may observe in a single campaign, making inference of the viewing geometry and bulk fluid properties of \s{} an appealing application of the model in the near term.

To conclude, we have found strong evidence of innate degeneracies in the black hole accretion flow system; purely equatorial emission is capable of reproducing images of more general morphologies with sometimes drastically different mass, spin, viewing inclination, and spin axis position angle. This result strongly motivates model fitting the most general emissivity distribution possible in order to infer black hole parameters with appropriate uncertainties in the absence of calibration to GRMHD simulations. Meanwhile, we have produced an emissivity model fitting approach that makes posteriors on black hole parameters achievable in a matter of hours on a personal computer; the approach generalizes to arbitrary combinations of optically thin emitting surfaces. Future work \citep{Chang_2022_in_prep} will expand the toy model to off-equatorial emission, enabling model fits much better suited to funnel-dominated emission of SANEs with large $R_{\rm high}$ (see, e.g., \citet{Wong_2021} for a recent examination of emission geometries in GRMHD).
We expect that the combination of inflowing disk surfaces and outflowing jet surfaces will provide a minimal, fast emissivity prescription that will successfully recover axisymmetric and quiescent properties of realistic accretion flows.

\begin{acknowledgements}
The authors would like to thank Ramesh Narayan, Elizabeth Himwich, and Maciek Wielgus for many useful discussions. We also thank Lindy Blackburn for computational resources and support. We thank our referee for a thoughtful review of our manuscript. This work was supported by the Black Hole Initiative at Harvard University, which is funded by grants from the John Templeton Foundation and the Gordon and Betty
Moore Foundation to Harvard University. This work was also supported by National Science Foundation grants AST 1935980 and AST 2034306 and the Gordon and Betty Moore Foundation (GBMF-5278).  A.C. acknowledges support from the Princeton Gravity Initiative. A.C. was supported by the NASA Hubble Fellowship grant HST-HF2-51431.001-A awarded by the Space Telescope Science Institute, which is operated by the Association of Universities for Research in Astronomy, Inc., for NASA, under contract NAS5-26555.

\end{acknowledgements}
\appendix
\numberwithin{equation}{section}
\section{Full Posteriors of the correctly specified self-fit}
\label{sec:fulln0n1}

\begin{figure*}
    \centering
    \includegraphics[width=\textwidth]{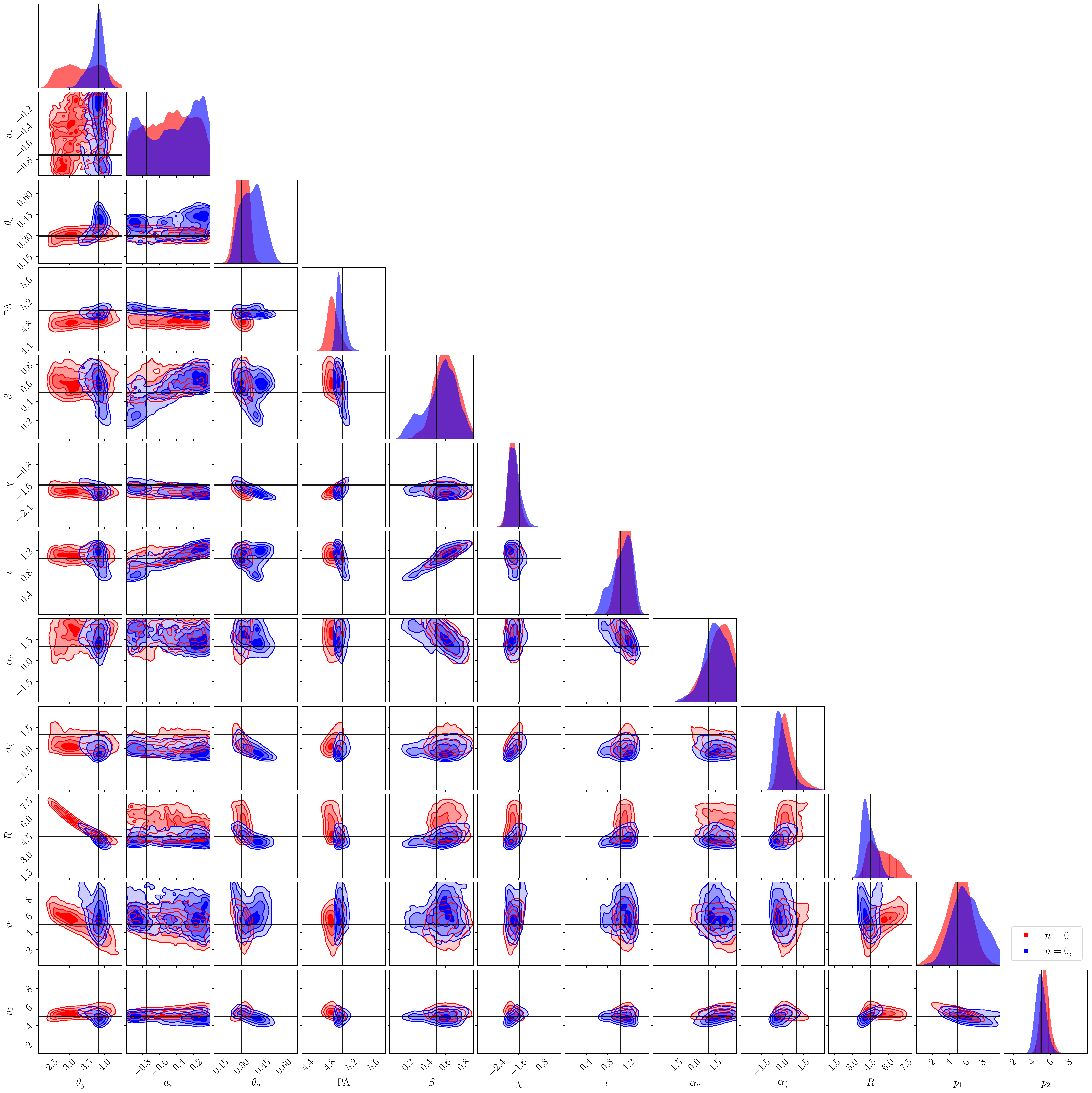}
    \caption{Same as \autoref{fig:n0_n1_selfcomp} but all fit parameters are shown.}
    \label{fig:n0_n1_selfcomp_all}
\end{figure*}

\autoref{fig:n0_n1_selfcomp_all} shows the full posteriors from the fit of both $n=0$ and $n=0,1$ models to themselves with identical sampling. Certain parameters, most notably the black hole spin $a_*$ and the emission profile shape parameters $p_1$ and $p_2$, are comparably unconstrained regardless of sub-image inclusion when fitting 2017 EHT \m{} data. The wide posterior in $a_*$ appears to arise from minor variation in $\iota$ and $\beta$, each of which have subtle effects on image asymmetries that increase with inclination, much like $a_*$.

\software{\texttt{eht-imaging} \citep{Chael_closure},
          \texttt{ipole} \citep{IPOLE_2018},
          \texttt{dynesty} \citep{Speagle_2020},
          \texttt{kgeo} \citep{Chael_kgeo_2022},
          Matplotlib \citep{matplotlib},
          Numpy \citep{numpy}}

\bibliography{main}

\end{document}